\documentclass[cits]{PoS}

\title{Topology, the Wilson flow and the HMC algorithm}
\ShortTitle{Topology, the Wilson flow and the HMC algorithm}

\author{\speaker{M.~L\"uscher}\\ 
        CERN, Physics Department, 1211 Geneva 23, Switzerland\\
        E-mail: \email{luscher@mail.cern.ch}}

\abstract{
An old and apparently persistent problem in numerical lattice QCD 
is that the simulations tend to get trapped in a sector
of fixed topological charge when the lattice spacing is taken to zero.
The effect sets in very rapidly and 
may invalidate the simulation results in certain cases.
In this talk, the issue is discussed using the Wilson flow
as a tool. The flow has a simple scaling behaviour
and allows one to understand how exactly the topological 
sectors emerge in the continuum limit. 
Further studies however suggest that the observed
slowdown of the simulations at small lattice spacings
is only partly caused by the emergence of the 
sectors.\\[0.5cm]
}

\FullConference{The XXVIII International Symposium on Lattice Field Theory, 
                Lattice2010\\
		June 14-19, 2010\\
		Villasimius, Italy}

\begin{document}





\def\blue{\color{blue}}
\def\darkblue{\color{darkblue}}
\def\black{\color{black}}
\def\magenta{\color{magenta}}
\def\red{\color{red}}
\def\darkred{\color{darkred}}
\def\redtwo{\color{redtwo}}
\def\white{\color{white}}
\def\lightgrey{\color{lightgrey}}


\def\rf#1{{\magenta\scriptsize #1}}


\def\rmd{{\rm d}}
\def\rmD{{\rm D}}
\def\rme{{\rm e}}
\def\rmO{{\rm O}}
\def\rmU{{\rm U}}


\def\rz{{\mathbb R}}
\def\gz{{\mathbb Z}}
\def\nz{{\mathbb N}}
\def\Im{{\rm Im}\,}
\def\Re{{\rm Re}\,}


\def\defeq{\mathrel{\mathop=^{\rm def}}}
\def\proof{\noindent{\sl Proof:}\kern0.6em}
\def\endproof{\hskip0.6em plus0.1em minus0.1em
\setbox0=\null\ht0=5.4pt\dp0=1pt\wd0=5.3pt
\vbox{\hrule height0.8pt
\hbox{\vrule width0.8pt\box0\vrule width0.8pt}
\hrule height0.8pt}}
\def\sfrac#1#2{\hbox{$\frac{#1}{#2}$}}
\def\dual{\mathstrut^*\kern-0.1em}
\def\mod{\;\hbox{\rm mod}\;}
\def\ring{\mathaccent"7017}
\def\slash#1{\setbox0=\hbox{$#1$}
    \setbox1=\hbox{$/$}
    #1\kern-\wd0\kern1.5pt/\kern-\wd1\kern\wd0}
\def\lvec#1{\setbox0=\hbox{$#1$}
    \setbox1=\hbox{$\scriptstyle\leftarrow$}
    #1\kern-\wd0\smash{
    \raise\ht0\hbox{$\raise1pt\hbox{$\scriptstyle\leftarrow$}$}}
    \kern-\wd1\kern\wd0}
\def\rvec#1{\setbox0=\hbox{$#1$}
    \setbox1=\hbox{$\scriptstyle\rightarrow$}
    #1\kern-\wd0\smash{
    \raise\ht0\hbox{$\raise1pt\hbox{$\scriptstyle\rightarrow$}$}}
    \kern-\wd1\kern\wd0}
\def\boxit#1{\vbox{\hrule height2pt\hbox{\vrule width2pt
    \kern10pt\vbox{\kern10pt#1\kern10pt}\kern10pt\vrule width2pt}
    \hrule height2pt}}
\def\Arw{{\redtwo\bm\Rightarrow}}


\def\nab#1{{\nabla\kern-1.5pt_{#1}\kern1.0pt}}
\def\nabstar#1{\nabla\kern0.5pt\smash{\raise5.0pt\hbox{$\ast$}}
               \kern-7.5pt_{#1}\kern2.0pt}
\def\drv#1{{\partial_{#1}}}
\def\drvstar#1{\partial\kern1.0pt\smash{\raise4.5pt\hbox{$\ast$}}
               \kern-6.5pt_{#1}\kern2pt}
\def\drvtilde#1#2{{\tilde{\partial}_{#1}^{#2}}}


\def\MeV{{\rm MeV}}
\def\GeV{{\rm GeV}}
\def\TeV{{\rm TeV}}
\def\fm{{\rm fm}}
\def\MSbar{\overline{\rm MS\kern-0.5pt}\kern0.5pt}


\def\euler{\gamma_{\rm E}}


\def\psibar{\overline{\psi}}
\def\chibar{\overline{\chi}}
\def\psitilde{\widetilde{\psi}}
\def\ubar{\bar{u}}
\def\dbar{\bar{d}}
\def\sbar{\bar{s}}
\def\Ubar{V}


\def\dirac#1{\gamma_{#1}}
\def\diracstar#1#2{
    \setbox0=\hbox{$\gamma$}\setbox1=\hbox{$\gamma_{#1}$}
    \gamma_{#1}\kern-\wd1\kern\wd0
    \smash{\raise4.5pt\hbox{$\scriptstyle#2$}}}
\def\dirachat{\hat{\gamma}_5}


\def\SUtwo{{\rm SU(2)}}
\def\SUthree{{\rm SU(3)}}
\def\SUn{{\rm SU}(N)}
\def\tr{{\rm tr}}
\def\Tr{{\rm Tr}}
\def\Ad{{\rm Ad}\,}
\def\Group{{\rm SU}(3)}
\def\Lie{\mathfrak{su}(3)}


\def\trans{{\cal F}}
\def\transs{\trans_{*}}
\def\transt{\trans_{t}}
\def\transts{\trans_{t,*}}
\def\transne{\trans_{n\eps}}
\def\transnes{\trans_{n\eps,*}}
\def\euler#1{{\cal E}_{#1}}
\def\du{\partial}


\def\Sw{S_{\rm w}}
\def\Sflow{\widetilde{S}}
\def\Wl{{\cal W}}
\def\Proj{{\cal P}}
\def\Nf{N_{\rm f}}


\def\Obs{{\cal O}}
\def\tauint{\tau_{\rm int}}
\def\ren#1{#1_{\rm R}}


\def\mom{\pi}
\def\eps{\epsilon}

\def\vsp{{\vphantom{$a_b$}}}
\def\thicktablerule{\hrule height1pt}
\def\thintablerule{\hrule height0.4pt}

\section{Introduction}

At present all widely used simulation algorithms for
lattice QCD with two or more flavours of sea quarks are based on
some version of the HMC algorithm \cite{HMC}
(see ref.~\cite{LesHouches} for a recent review).
The computer time required for such simulations  
depends on the time needed for an update step,
the integrated autocorrelation times of
the observables of interest and the desired statistics.
As a function of the lattice spacing $a$, 
and if the algorithm is implemented as usual,
the first factor scales roughly like $a^{-5}$,
but the autocorrelation times tend to be
difficult to determine and are often not known reliably 
if they are not small.

Large autocorrelation times are typically
observed in the case of quantities related to the
topological charge of the gauge field. While the definition of
the charge on the lattice is ambiguous to some extent, experience suggests
that the autocorrelation times are largely insensitive
to the exact choices one makes. For illustration 
a time series of charge measurements
is plotted in fig.~\ref{fig1}.
The presence of strong autocorrelations is evident
in this example, particularly so in the shaded region,
where the charge never changes sign and instead
oscillates around a value of $-10$. Note that
most QCD simulations published to date are
much shorter than the run shown in the figure.

The integrated autocorrelation times of physical quantities tend 
to grow when the continuum limit is approached.
Depending on the theory, the algorithm and the observable,
the asymptotic scaling behaviour can be very different,
but is normally power-like in the inverse lattice spacing 
with an exponent $z\leq 2$.
Del Debbio, Panagopoulos and Vicari \cite{DelDebbioTauQ}
however found some time ago that
the topological charge in the SU(3) gauge theory
has a much more rapidly (perhaps exponentially) increasing
autocorrelation time if the standard link-update algorithms
are used.
Further studies by Schaefer, Sommer and Virotta \cite{SchaeferTauQ}
later showed that the HMC algorithm is similarly inefficient
and there is ample evidence that the problem persists when the sea quarks 
are added to the theory \cite{SchaeferTauQ,RBC-UKQCD,MILC}.

In this talk, a few steps are taken towards a better
understanding of the dynamics of the HMC algorithm
and the mechanism that leads to the dramatic
slowdown of the simulations at small lattice spacings.
Issues to be addressed
are how exactly the topological sectors emerge in the continuum limit,
what the slow modes of the gauge field might be and whether perhaps
there is a simple way out. The discussion is largely based
on a new tool, the Wilson flow \cite{TrivMaps,WilsonFlow},
which is of some interest in its own right.

\section{Autocorrelations}

\subsection{Scaling behaviour}

Systematic scaling studies of autocorrelation times 
in lattice QCD have so far been limited to the case 
of the pure gauge theory \cite{DelDebbioTauQ,SchaeferTauQ}.
In all these studies, the autocorrelation time
of the topological charge turned out to increase at least like $a^{-5}$
at lattice spacings $a$ below $0.1$ fm or so.
The algorithms considered (the HMC, the
DD-HMC and the well-known link-update algorithms)
appear to behave similarly in this respect.
In particular, once the link-visiting
frequency is divided out,
the block size used in the DD-HMC algorithm
does not have a significant influence on the 
autocorrelation times \cite{SchaeferTauQ}.

The onset of the rapid growth of the autocorrelation time of
the topological charge, and thus its value at a given
lattice spacing, however depends 
on the chosen lattice action and the simulation algorithm.
Increasing the length of the molecular-dynamics
trajectories in the HMC algorithm can be
beneficial \cite{SchaeferTauQ}, for example, while the addition of 
six-link terms to the Wilson plaquette action 
may have an adverse effect \cite{SchaeferTauQ,RBC-UKQCD}. 

When the sea quarks are included in the simulations, the 
situation becomes considerably more complicated,
because the autocorrelations may now also depend on the
number of quark flavours, the quark masses
and chosen the fermion action 
\cite{SchaeferTauQ,RBC-UKQCD,MILC}.
These dependencies and the one on the lattice spacing 
remain to be studied in detail,
but the experience made so far (which is
sometimes only based on a visual inspection of measurement histories)
shows that the autocorrelation time of the 
topological charge is again very rapidly growing when 
the lattice spacing is reduced from $0.1$ fm to $0.05$ fm or
even smaller values.

If the leading exponential autocorrelation time is assumed
to grow proportionally to $a^{-5}$,
the total computational effort required for HMC simulations 
of QCD is expected to scale like $a^{-10}$ at fixed physics, i.e.~the 
cost of the simulations increases by about three orders of magnitude
when the lattice spacing is divided by $2$.
This estimate may be a bit pessimistic, but 
it is quite clear that the simulations required for
safe extrapolations to the continuum limit 
are extremely challenging.
Algorithmic improvements or viable ways of 
bypassing the slowing down of the simulations are certainly
highly desirable at this point.

\subsection{Autocorrelation effects in short runs}

A question often asked in this context is whether 
the simulations really need to be very much longer than 
the leading exponential autocorrelation time.
In particular, if the quantities of interest
are only weakly coupled to the slow modes of 
the algorithm, 
the results obtained in shorter 
runs may conceivably be correct within 
statistical errors.

\begin{figure}
\centering
\includegraphics[width=.6\textwidth,clip]{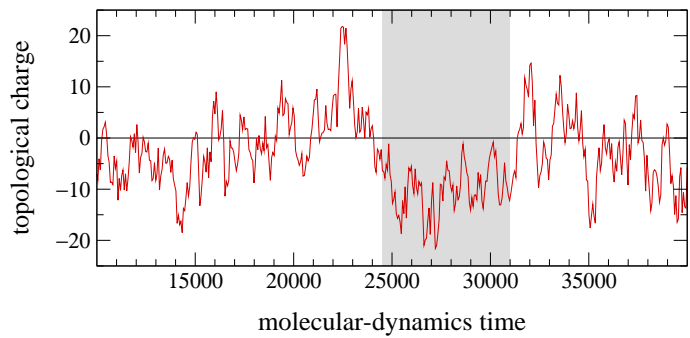}
\caption{History of the topological charge in a simulation of the
pure SU(3) gauge theory on a $64\times32^3$ lattice with spacing
$a=0.07$ fm. The plain HMC algorithm was
used in this test run, with a trajectory length of $2$ units of 
molecular-dynamics time and an acceptance rate of $83\%$.}
\label{fig1}
\end{figure}

In general, the expectation values calculated in such 
short runs must be expected to be biased to some extent.
Considering again the measurement history plotted in
fig.~\ref{fig1}, for example, it is clear that the topological charge is
incorrectly sampled in this case if
runs not very much longer than $5000$ molecular-dynamics
time units are performed. 
The expectation values 
of most observables are then affected
by terms inversely proportional to the space-time volume
\cite{VolumeEffectsI,VolumeEffectsII}. 
Physical quantities like the masses of the $\eta$ and $\eta'$
mesons, but also hadronic matrix elements of 
pseudo-scalar densities are likely to be
strongly sensitive to these effects.

When arguing for short runs, one would need to provide
a practical procedure that allows the effects of the slow modes
on the calculated expectation values to be estimated. 
Whether they are in fact negligible 
in the cases of interest is otherwise difficult to tell
and one is left with results
that may or may not be correct.

\begin{figure}
\centering
\includegraphics[width=.5\textwidth,clip]{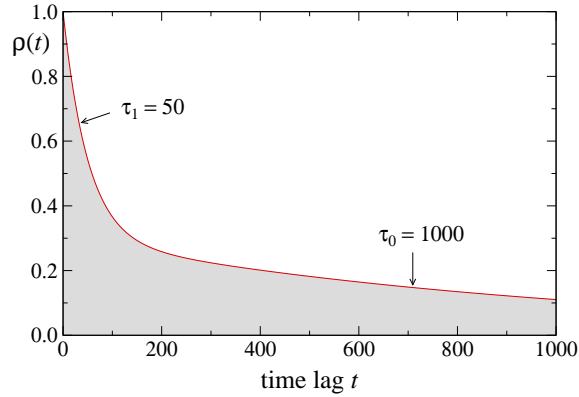}
\caption{Example of a normalized autocorrelation function 
$\rho(t)$ that is dominated by two modes with exponential 
autocorrelation times $\tau_0\gg\tau_1$.}
\label{fig2}
\end{figure}

\subsection{Statistical error estimation in presence of slow modes}

Large exponential autocorrelation times can also lead to 
an underestimation of integrated autocorrelation times
and thus of the associated statistical errors.
To illustrate this point, consider a normalized autocorrelation 
function
\begin{equation}
  \rho(t)=|c_0|^2\rme^{-t/\tau_0}+|c_1|^2\rme^{-t/\tau_1}+\ldots
  \label{rho}
\end{equation}
which is dominated by two eigenmodes of the
simulation transition probability, a slow mode
with exponential autocorrelation time $\tau_0$ and 
a fast mode with autocorrelation time $\tau_1\ll\tau_0$
(see fig.~\ref{fig2}).
The coefficients $|c_0|^2$ and $|c_1|^2$ in this formula
measure how strongly these modes couple to the observable
considered. 
Note that the integrated autocorrelation time
\begin{equation}
  \tauint\simeq\tau_0|c_0|^2+\tau_1|c_1|^2+\ldots
  \label{tauint}
\end{equation}
is equal to the area under the curve shown in the figure. 
Whether $\tauint$ 
can be easily estimated now depends on whether the area under the tail
of the curve dominates or not. If it does not, i.e.~if 
$\tau_0|c_0|^2\ll\tau_1|c_1|^2$,
the relevant exponential autocorrelation time is $\tau_1$ and 
the statistical error is correctly obtained from
simulations a few hundred times longer than $\tau_1$.
Runs very much longer than $\tau_0$ are however 
required in the other case to be able to control 
the situation.

Note that the contribution of the slow mode to the 
statistical error is not guaranteed to be negligible 
even if the coupling $|c_0|^2$ is very small.
These cases are actually particularly
difficult to treat correctly, because
the autocorrelation function in the tail
is very small, while the area under the tail may not be so.
Since the autocorrelation function itself 
can only be calculated up to some statistical uncertainty,
it may then be quite impossible to exclude this case,
except when $\tau_0$ is known or can
at least be bounded from above \cite{SchaeferTauQ}.

\section{Wilson flow}

The high-frequency components of the gauge field 
are usually weakly coupled to the slow modes of the HMC algorithm
and tend to be efficiently updated.
What exactly one means by the smooth components of the 
field is not entirely clear, however, and even less so
which of their properties (apart from the topological charge)
are slowly sampled by the algorithm. 
As explained in the following, 
the Wilson flow \cite{TrivMaps,WilsonFlow}
is a renormalizable smoothing operation
that allows these questions to be addressed on
theoretically solid ground.

\subsection{Flow equation}

\begin{figure}
\centering
\includegraphics[width=.275\textwidth,clip]{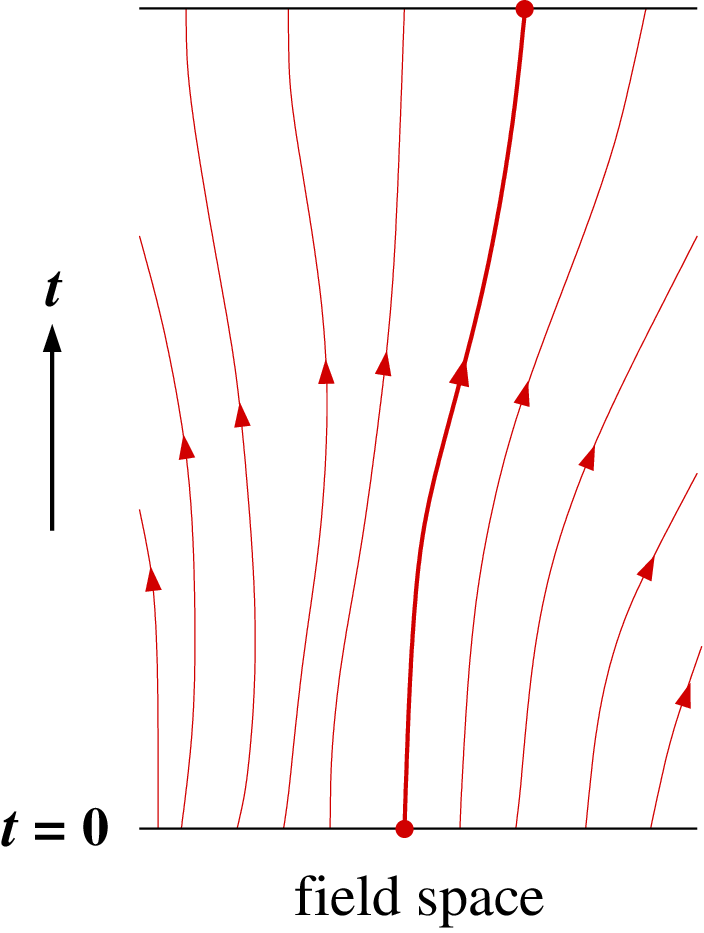}
\caption{
Equation (3.1) generates a flow in the space of 
lattice gauge fields. On a finite lattice, 
the existence, smoothness and uniqueness of the flow is 
rigorously guaranteed at all flow times \cite{TrivMaps}. Moreover, 
at any given time $t$,
the mapping $U\to V_t$ is an invertible 
transformation of field space, because the flow can be integrated
both forward and backward in time.
}
\label{fig3}
\end{figure}

For any given lattice gauge field $U(x,\mu)$,
the first-order differential equation
\begin{equation}
  \dot{V}_t(x,\mu)=-g_0^2\left\{\partial_{x,\mu}\Sw(V_t)\right\}V_t(x,\mu),
  \qquad
  \left.V_t(x,\mu)\right|_{t=0}=U(x,\mu),
  \label{Wflow}
\end{equation}
defines a trajectory $V_t(x,\mu)$ of fields
parameterized by the ``flow time'' $t$ (see fig.~\ref{fig3}; differentiation
with respect to $t$ is abbreviated by a dot).
In this equation, $\Sw(V_t)$ denotes the Wilson plaquette action
\cite{Wilson} of the field $V_t$ at gauge coupling $g_0$ 
and $\partial_{x,\mu}\Sw(V_t)$ its (Lie algebra valued)
variation with respect to the link variable $V_t(x,\mu)$.
Note that the coupling cancels in the flow equation.

Along the Wilson flow,
the plaquette action decreases monotonically,
$\dot{S}_{\rm w}\leq0$, and the gauge field
tends to become smoother.
The flow is in fact generated by infinitesimal ``stout''
link-smearing steps \cite{Stout} and thus shares
some properties with this popular smoothing procedure.
Eventually the flow drives
the field towards the stationary points of the action,
but contrary to what may be assumed, the large-time
regime is a highly non-perturbative one in QCD. In particular,
it may not be meaningful to study the large-time behaviour
of the flow separately from
the continuum limit of the theory.

\subsection{QED --- a soluble case}

In the continuum limit, the flow equation (\ref{Wflow}) assumes the form
\begin{equation}
  \dot{B}_{\mu}=D_{\nu}G_{\nu\mu},
  \qquad
  \left.B_{\mu}\right|_{t=0}=A_{\mu},
  \label{Bflow}
\end{equation}
where $A_{\mu}(x)$ is the fundamental gauge field, $B_{\mu}(t,x)$ the
time-dependent gauge field and $G_{\mu\nu}(t,x)$ the associated
field tensor. In QED
the equation is a linear diffusion equation,
whose solution is given by 
\begin{equation}
  B_{\mu}(t,x)=\int\rmd^4y\,K_t(x-y)A_{\mu}(y)+\hbox{gauge terms},
  \qquad
  K_t(z)=\frac{\rme^{-\frac{z^2}{4t}}}{(4\pi t)^2}.
  \label{BflowQED}
\end{equation}
The flow thus averages the gauge field over a
spherical range with mean-square radius equal to $\sqrt{8t}$.
Note that, as is already clear from eq.~(\ref{Bflow}), 
the flow time $t$ has engineering
dimension $[\hbox{length}]^2$.

Since the field generated by the flow is linearly related to the 
fundamental gauge field, its correlation functions
\begin{eqnarray}
  \langle B_{\mu_1}(t,x_1)\ldots B_{\mu_n}(t,x_n)\rangle&=&
  e_0^n\int\rmd^4y_1\ldots\rmd^4y_n\,K_t(x_1-y_1)\ldots K_t(x_n-y_n)
  \nonumber\\[1.5ex]
  &&
  \qquad\times G_0(y_1,\ldots,y_n)_{\mu_1\ldots\mu_n}+
  \hbox{gauge terms}
  \label{Bcorr}
\end{eqnarray}
are proportional to the (full, bare) photon $n$-point functions
$G_0(y_1,\ldots,y_n)_{\mu_1\ldots\mu_n}$. A power of the bare 
electron charge $e_0$
appears in this equation, because the canonically normalized 
photon field is $e_0A_{\mu}$ rather than $A_{\mu}$
(which is normalized so that the covariant derivatives do not
involve the charge). Note also that the heat kernels $K_t(x_k-y_k)$
play the r\^ole of smooth test functions in this formula. In 
particular, the renormalization of the 
correlation function is achieved simply by
renormalizing the bare charge and the photon $n$-point function 
according to
\begin{equation}
  e_0=Z_3^{-1/2}\ren{e},
  \qquad
  G_0=Z_3^{n/2}\ren{G}.
\end{equation}
The fact that both $e_0$ and $G_0$ renormalize with the
same renormalization constant $Z_3$ is a consequence
of the gauge Ward identity in this theory
(a gauge-invariant regularization is assumed here).
Since $e_0^nG_0=\ren{e}^n\ren{G}$, this 
shows that the field $B_{\mu}(t,x)$ is, at all positive 
flow times $t$
and up to its gauge degrees of freedom,
a \emph{renormalized smooth gauge field}. 

Provided the bare charge is expressed through the renormalized one,
the correlation functions of the field tensor $G_{\mu\nu}$
thus do not require any renormalization and converge to
well-defined smooth functions of the space-time coordinates
when the regularization of the theory is removed. Moreover,
since QED is asymptotically free at low energies, the 
behaviour of the correlation functions at large flow times
is described by leading-order perturbation theory.
A short calculation then leads to the formula
\begin{equation}
  \lim_{t\to\infty}\{t^2\langle G_{\mu\nu}G_{\mu\nu}\rangle\}
  =\frac{3\ren{e}^2}{32\pi^2},
\end{equation}
which shows that the field obtained by 
the Wilson flow contains some interesting physical 
information.

\subsection{Properties of the Wilson flow in QCD}

In QCD the flow equation is non-linear and the renormalization of the 
theory is much more complicated than in QED. Whether
the Wilson flow generates a renormalized gauge field is however
a question that can be studied in perturbation
theory. In particular,
the expectation value
of the gauge-invariant density
\begin{equation}
  E=-\sfrac{1}{2}\tr\{G_{\mu\nu}G_{\mu\nu}\}
\end{equation}
can be easily worked out
to next-to-leading order in the gauge coupling.
Dimensional regularization may be used in this calculation and
it then turns out that $\langle E\rangle$ does not require renormalization
to this order, i.e.~the divergent terms are all canceled by the 
renormalization of the coupling. In the $\MSbar$ scheme,
the one-loop formula obtained in this way is \cite{WilsonFlow}
\begin{eqnarray}
  \langle E\rangle&=&\frac{3}{4\pi t^2}\alpha(q)
  \left\{1+k_1\alpha(q)+\ldots\right\},
  \qquad
  q=(8t)^{-1/2},
  \label{PTexpansion}
  \\[2.0ex]
  k_1&=&1.0978+0.0075\times\Nf,
\end{eqnarray}
where $\Nf$ denotes the number of massless sea quarks 
and the running coupling 
$\alpha(q)$ is evaluated at a momentum scale $q$ 
equal to the inverse of the 
leading-order smoothing range of the flow.

\begin{figure}
\centering
\includegraphics[width=.6\textwidth,clip]{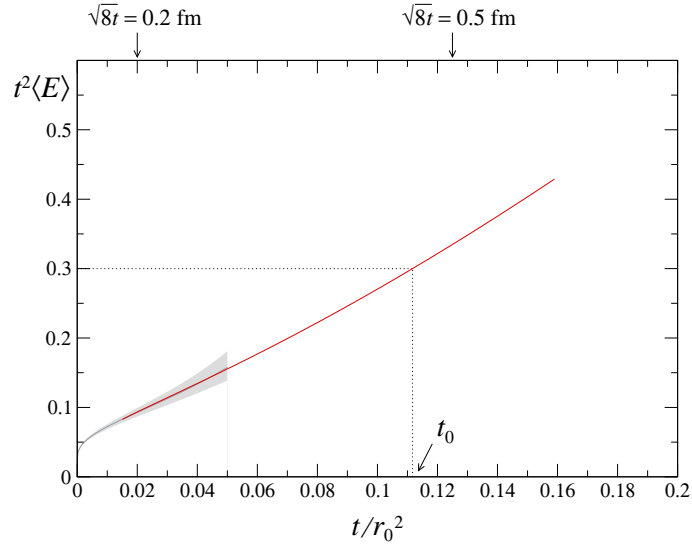}
\caption{
Plot of $t^2\langle E\rangle$ in the pure SU(3) gauge
theory as a function of the flow time $t$ given in units of the Sommer
scale $r_0$ \cite{SommerScale,GuagnelliEtAl}.
The red solid line was obtained
through numerical simulation of
a $96\times48^3$ lattice with spacing $a=0.05$ fm. One-loop
perturbation theory [eq.~(3.8)]
and the known value of the $\Lambda$-parameter \cite{LambdaParm}
yield the grey curve with an error margin (grey area)
deriving from the error on $r_0\Lambda$.
}
\label{fig4}
\end{figure}

Beyond perturbation theory, the expectation value of $E$ can be 
computed straightforwardly using the lattice formulation of the 
theory and numerical simulations. An accurate numerical integration
of the flow equation (\ref{Wflow}) is required in these calculations,
but the computer time needed for the integration is negligible
in practice if a suitable higher-order integrator is used \cite{WilsonFlow}. 
The result of a computation along these lines is 
plotted in fig.~\ref{fig4} in a range of the flow time
corresponding to smoothing ranges from about $0.2$ to $0.5$ fm
(the statistical errors are not visible on the scale of the plot). 
Since QCD is asymptotically free, 
the perturbation expansion (\ref{PTexpansion}) 
only applies at small flow times,
but as can be seen from the curves shown in 
fig.~\ref{fig4}, the transition from the small-time to the 
non-perturbative regime is very smooth.
Note that $t^2\langle E\rangle$ increases roughly
linearly with $t$ in the non-perturbative regime, 
at least so in the range shown in the plot,
a behaviour which is completely different from 
the one in QED.

\begin{figure}
\centering
\includegraphics[width=.6\textwidth,clip]{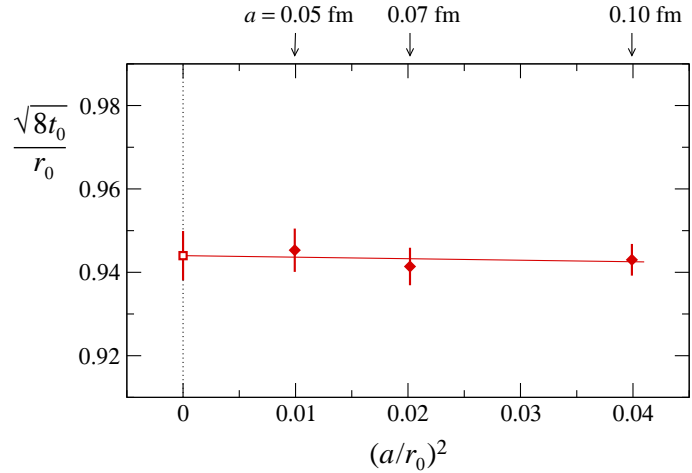}
\caption{
Scaling behaviour of the reference time $t_0$ 
in the pure SU(3) gauge theory. The data (full points) were
obtained on lattices of size $48\times24^3$, $64\times32^3$
and $96\times48^3$ with spacings approximately equal to 
$0.1$, $0.07$ and $0.05$ fm, respectively.
The continuum limit (open point) is reached 
through a linear fit in $a^2$ of the three data points (solid line).
}
\label{fig5}
\end{figure}

If $\langle E\rangle$ does not require renormalization, one
expects $t^2 \langle E\rangle$ to be a universal function of 
$t/r_0^2$ up to lattice effects that vanish proportionally to 
a positive power of the lattice spacing.
The scaling can be checked by introducing a reference time
$t_0$ through the implicit equation
\begin{equation}
  \left.t^2\langle E\rangle\right|_{t=t_0}=0.3
  \label{RefTime}
\end{equation}
(see fig.~\ref{fig4}). Simulations at three values of 
the lattice spacings then 
show that the dimensionless ratio $\sqrt{8t_0}/r_0$ does in fact smoothly
converge to the continuum limit,
the lattice-spacing effects being less than a percent
in the range covered by the simulations (fig.~\ref{fig5}).
Little doubt thus remains that the Wilson flow maps the gauge
field to a renormalized smooth field as in QED.

\section{Topological sectors}

The question of how exactly the topological sectors emerge in 
lattice QCD can now be answered by performing the field
transformation $U\to V=V_{t_0}$ in the functional integral,
$t_0$ being the reference flow time introduced in the 
previous section. The smoothing range is about $0.5$ fm
in this case, i.e.~on average the fluctuations 
of the gauge field with wavelengths up to the confinement
radius are smoothed out. Somewhat surprisingly,
the Jacobian of the transformation can be 
analytically expressed through the Wilson action
\cite{TrivMaps}. The expectation value of any 
observable $\Obs(U)$ then assumes the form
\begin{equation}
  \langle\Obs\rangle={1\over{\cal Z}}
  \int\rmD[V]\,\Obs(U)\,\rme^{-\tilde{S}(V)},
  \label{TransInt}
\end{equation}
where the action is given by
\begin{equation}
  \tilde{S}(V)=S(U)+\frac{16\kern0.5pt g_0^2}{3\kern0.5pt a^2}
  \int_0^{t_0}\rmd t\,\Sw(V_t).
  \label{NewAction}
\end{equation}
In this formula, the fields $U$ and $V_t$ 
are considered to be functions of $V$ and
$S(U)$ denotes the action of the theory (including the 
quark determinants if any)
before the transformation.

\begin{figure}
\centering
\includegraphics[width=.65\textwidth,clip]{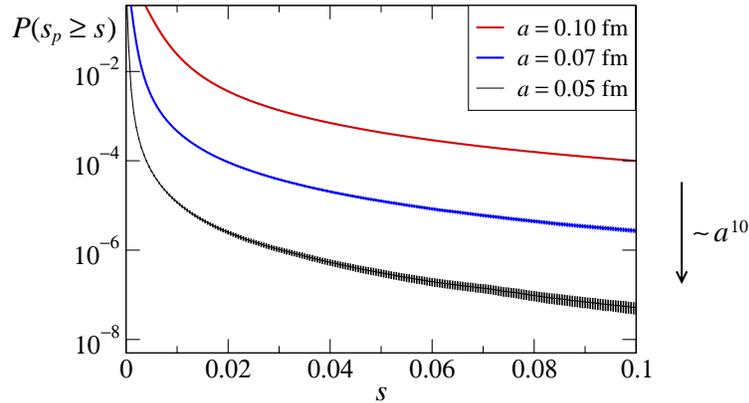}
\caption{
Probability for the value of $s_p$ [eq.~(4.3)] on a given plaquette $p$
to be above a specified threshold~$s$.  
The three curves were obtained 
in the pure SU(3) gauge theory with Wilson 
plaquette action, using the same representative 
ensembles of gauge-field configurations as in the scaling test 
reported in fig.~5. 
}
\label{fig6}
\end{figure}

Both terms in the action (\ref{NewAction}) tend to suppress large
values of the plaquette observable
\begin{equation}
  s_p=\Re\tr\{1-V(p)\}
\end{equation}
(where $V(p)$ is the product of the link variables $V(x,\mu)$
around the plaquette $p$)
and thus force the plaquette loops to be close to unity.
Since $s_p$ is a field of dimension $4$, representing
the square of the gauge-field tensor in the plaquette plane, 
its expectation value is in fact expected to scale like $a^4$
in the continuum limit. Numerical studies
confirm this behaviour and show that large values
of $s_p$ are indeed very strongly suppressed
(see fig.~\ref{fig6}).

Many years ago, a theorem was established \cite{TopCharge,PhillipsStone}
which states that the space of all lattice gauge fields 
satisfying a certain smoothness condition decomposes
into topological charge sectors very much
like the space of continuous fields in the classical continuum theory.
In particular, the theorem applies
to the subspace of SU(3) gauge fields $V$ which satisfy
\begin{equation}
  s_p<0.067\hspace{0.5em}\hbox{for all plaquettes $p$}.
\end{equation}
According to the empirical results reported in
fig.~\ref{fig6}, and since the number of 
plaquettes in a fixed physical volume grows proportionally to 
$a^{-4}$, the fields that do \emph{not} fulfill this condition
occur with a probability of order $a^6$. 
In the functional integral (\ref{TransInt}),
the weight of these configurations (and thus of the region 
of field space, where the assignment of the
topological charge is ambiguous) is therefore rapidly 
decreasing when the lattice spacing is taken to zero.
In lattice gauge theory,
the emergence of the topological sectors
is thus seen to be a dynamical phenomenon, which sets in
close to the continuum limit.

\section{Slow modes of the HMC algorithm}

The field transformation considered in the previous section maps
the molecular-dynamics trajectories generated by the HMC algorithm
to trajectories of the transformed field $V=V_{t_0}$.
At lattice spacings where the regions of field
space ``between the topological sectors'' are strongly
suppressed, it is unlikely that such a trajectory
leads from one sector to another, because the
fields along the trajectories are, to a good approximation,
distributed according to their weight in the functional integral.
The emergence of the sectors and the freezing of the topological
charge in HMC simulations are thus directly related to each other.

\subsection{Do the smoothed fields move slowly?}

The algorithm may however slow down for other reasons
as well and there is no guarantee that the
gauge field is efficiently updated in a fixed sector.
In particular, the smooth fields obtained by the 
Wilson flow at flow time $t>0$ may conceivably be slowly moved
through configuration space, in which case
the autocorrelation times of local observables
constructed from these fields (such as $E$)
are expected to be large.

\begin{table}
\centering
\newdimen\digitwidth
\setbox0=\hbox{\rm 0}
\digitwidth=\wd0
\catcode`@=\active
\def@{\kern\digitwidth}
\begin{tabular}{cccc}
$t/t_0$ &
$\tauint[Q]$ &
$\tauint[Q^2]$ &
$\tauint[E]$ \\[0.0ex]
\hline\hline
\noalign{\vskip0.5ex}
   $0.18$ &
   $@65(5)@$ &
   $@30(2)@$ &
   $25(2)$ \\
   $0.35$ &
   $@67(5)@$ &
   $@32(2)@$ &
   $35(2)$ \\
   $0.70$ &
   $@68(6)@$ &
   $@33(2)@$ &
   $44(3)$ \\
\noalign{\vskip0.5ex}
\hline\hline
\end{tabular}
\hspace{1.0cm}
\begin{tabular}{cccc}
$t/t_0$ &
$\tauint[Q]$ &
$\tauint[Q^2]$ &
$\tauint[E]$ \\[0.0ex]
\hline\hline
\noalign{\vskip0.5ex}
   $0.18$ &
   $614(90)$ &
   $284(34)$ &
   $55(4)$ \\
   $0.36$ &
   $615(90)$ &
   $286(34)$ &
   $69(5)$ \\
   $0.72$ &
   $615(90)$ &
   $286(34)$ &
   $86(6)$ \\
\noalign{\vskip0.5ex}
\hline\hline
\end{tabular}
\caption{Integrated autocorrelation times in the pure SU(3)
gauge theory with Wilson action, measured on a 
$48\times24^3$ lattice with spacing $a=0.1$ fm (left table)
and on a $64\times32^3$ lattice with spacing $a=0.07$ fm (right table).
In both cases, the HMC algorithm with a trajectory length of $2$
and an acceptance rate of $83\%$ was used. All autocorrelation times 
and the trajectory length are given in units of molecular-dynamics
time.}
\label{tab1}
\end{table}

The measured autocorrelation times of $E$ listed in table~1
are actually monotonically increasing with the flow time $t$
and reach values up to $10$ times the 
autocorrelation time of the usual plaquette observable at $t=0$.
On the other hand, when the lattice spacing is reduced,
they do not appear to grow as rapidly as the autocorrelation times
of the topological charge $Q$ (in these studies,
$E$ and the charge density were both defined using 
the same symmetric plaquette-loop expression for the 
gauge-field tensor at the specified flow time).
The relatively slow updating of 
the smoothed field and the charge freezing are thus
two different effects, where one is not obviously driving
the other.

\subsection{Open boundary conditions}

One may be inclined to conclude at this point that 
the emergence of the topological sectors at small lattice 
spacings is the principal and perhaps only cause for 
the presence of very rapidly growing autocorrelation 
times in HMC simulations.
However, when choosing
open instead of periodic boundary conditions in the physical 
time direction, the topological sectors disappear and the
space of smooth fields becomes connected. 
One thus expects to observe only moderately increasing
autocorrelation times in this case if the 
slowdown of the algorithm is indeed mainly caused by the
separation of the sectors.

A lattice with time extent $T$ and open boundary conditions in time
does not wrap around in this direction, i.e.~there are no terms
in the action which couple the field variables at time
$x_0=0$ to those at the largest time $x_0=T-a$. In a
simulation program for periodic lattices,
these boundary conditions can often
be implemented simply by setting the time-like link variables 
at the largest time to zero. Note that open boundary conditions
preserve the gauge symmetry. In the continuum theory, they amount
to imposing Neuman boundary conditions,
\begin{equation}
  \left.F_{0k}(x)\right|_{x_0=0}=
  \left.F_{0k}(x)\right|_{x_0=T}=0,
  \qquad k=1,2,3,
\end{equation}
on the gauge field and homogeneous Dirichlet (Schr\"odinger functional)
boundary conditions on the quark fields, $F_{\mu\nu}$ being
the field tensor of the fundamental gauge field.
The field space becomes topologically trivial
when these boundary conditions are chosen.
In particular, instantons
can be smoothly moved in and out of the volume
through the boundaries at $x_0=0,T$.

The figures listed in table~\ref{tab2} show that the 
situation is in fact somewhat improved when passing from 
periodic to open boundary conditions (in both cases,
$Q$ denotes the sum of the topological charge density 
over all lattice points). As a function of the lattice
spacing, the autocorrelation time of $Q$ is 
however rapidly growing independently of the
chosen boundary condition. 
On the periodic lattices,
the quantization of the topological charge
thus appears to be only one of several causes 
of the slowdown of the HMC algorithm.

\begin{table}
\centering
\newdimen\digitwidth
\setbox0=\hbox{\rm 0}
\digitwidth=\wd0
\catcode`@=\active
\def@{\kern\digitwidth}
\begin{tabular}{lccc}
&
$\tauint[Q]$ &
$\tauint[Q^2]$ &
$\tauint[E]$ \\[0.0ex]
\hline\hline
\noalign{\vskip0.5ex}
   periodic&
   $@68(6)@$ &
   $@33(2)@$ &
   $44(3)$ \\
   open&
   $@61(6)@$ &
   $@27(2)@$ &
   $38(3)$ \\
\noalign{\vskip0.5ex}
\hline\hline
\end{tabular}
\hspace{1.0cm}
\begin{tabular}{lccc}
&
$\tauint[Q]$ &
$\tauint[Q^2]$ &
$\tauint[E]$ \\[0.0ex]
\hline\hline
\noalign{\vskip0.5ex}
   periodic &
   $615(90)$ &
   $286(34)$ &
   $86(6)$ \\
   open &
   $384(56)$ &
   $155(20)$ &
   $77(6)$ \\
\noalign{\vskip0.5ex}
\hline\hline
\end{tabular}
\caption{Comparison of autocorrelation times
measured on lattices with periodic and open boundary conditions 
in the physical time direction. The lattices are the same as in table~1
(left: $a=0.1$ fm, right: $a=0.07$ fm), the HMC parameters
are also the same and the flow time $t$
is equal to $0.7\times t_0$ in all cases.}
\label{tab2}
\end{table}

\subsection{Instantons and the chiral limit}

One of the distinguishing features of non-abelian gauge theories
is the existence of the instanton solutions of 
the classical field equations. The fact that
the action has a large manifold of nearly
stationary points (the configurations built from many distant instantons and 
anti-instantons)
may conceivably play an important dynamical r\^ole in 
these theories, but so far it proved to be difficult 
to actually show this.
In the present context, the observation is perhaps
of some relevance, because the fields generated by the Wilson flow 
at large flow times tend to be close to the stationary points
of the action.

In QCD with light sea quarks, 
the fluctuations of the topological charge are suppressed
and the smooth fields obtained by the Wilson
flow may therefore be characteristically different from 
the ones in the pure gauge theory.
The slow modes of the HMC algorithm consequently
need not be the same and the autocorrelation times 
of the usual quantities may turn out to 
have a significant dependence on the quark masses.
However, as already mentioned, systematic 
studies of autocorrelation times 
await to be performed in QCD with dynamical quarks.

\section{Conclusions}

Numerical lattice QCD rests on the assumption 
that simulations at sufficiently small lattice spacings
can eventually be performed to be able to control the
discretization errors. It is therefore of central importance
to overcome the poor scaling behaviour of the currently available
simulation algorithms. 
Finding better algorithms is difficult, however, 
and probably requires a more detailed understanding 
of why the HMC algorithm slows down when the lattice
spacing is taken to zero.
 
The Wilson flow is a useful tool in this context, because
it allows the properties of the gauge field at different
length scales to be studied consistently with the 
renormalization of the theory. 
In particular, the way in which the topological sectors
emerge close to the continuum limit is made transparent
through the flow. The emergence of these sectors is
certainly one of the principal causes of the slowdown
of the HMC algorithm, but the fact that the situation
improves only slightly on lattices with open boundary conditions
is a bit surprising and remains unexplained.
It seems safe to conclude, however, 
that there are important further sources of inefficiency
and that the problem is likely to persist at fixed topological charge
(the rapid slowdown of the algorithm is in this case 
expected to be revealed
when considering the sum of the charge density over half the lattice).

Choosing open boundary conditions in the physical time direction
is nevertheless an interesting option, because the barriers
between the sectors disappear in this case, while the transfer
matrix (and thus much of the physics) is the same as with
periodic boundary conditions. From the point of view of the 
simulation algorithms, the absence of the barriers 
represents an important simplification. Tunneling transitions
are then not required anymore and one is left with the task
of finding an algorithm that moves
the field $V_t$ (at, say, $t=t_0$) efficiently
through the space of smooth configurations.

\acknowledgments

I am indebted to Filippo Palombi, Stefan Schaefer and Rainer Sommer 
for useful discussions on various issues related to this talk.
All numerical simulations reported here were performed on a
dedicated PC cluster at CERN. I am grateful
to the CERN management for providing the required funds
and to the CERN IT Department for technical support.


\begin{thebibliography}{99}


\bibitem{HMC}
S.~Duane, A.~D.~Kennedy, B.~J.~Pendleton, D.~Roweth,
\emph{Hybrid Monte Carlo},
\emph{Phys.~Lett.}~{\bf B195} (1987) 216.



\bibitem{LesHouches}
M.~L\"uscher,
\emph{Computational strategies in lattice QCD},
Lectures given at the
Summer School on ``Modern perspectives in lattice QCD'',
Les Houches, August 3-28, 2009,
arXiv:1002.4232 [hep-lat].



\bibitem{DelDebbioTauQ}
L.~Del Debbio, H.~Panagopoulos, E.~Vicari,
\emph{$\theta$-dependence of SU(N) gauge theories},
\emph{JHEP} {\bf 08} (2002) 044.



\bibitem{SchaeferTauQ}
S.~Schaefer, R.~Sommer, F.~Virotta,
\emph{Investigating the critical slowing down of QCD simulations},
PoS (LAT2009) 032;
\emph{Critical slowing down and error analysis in lattice QCD simulations},
arXiv:1009.5228 [hep-lat].

\bibitem{RBC-UKQCD}
D.~J.~Antonio et al.~(RBC and UKQCD Collab.),
\emph{Localization and chiral symmetry in 2+1 flavor domain wall QCD},
\emph{Phys.~Rev.~}{\bf D77} (2008) 014509.

\bibitem{MILC}
A.~Bazavov et al.~(MILC Collab.),
\emph{Topological susceptibility with the asqtad action},
\emph{Phys.~Rev.~}{\bf D81} (2010) 114501.



\bibitem{TrivMaps}
M. L\"uscher,
\emph{Trivializing maps, the Wilson flow and the HMC algorithm},
\emph{Commun.~Math.~Phys.}~{\bf 293} (2010) 899.



\bibitem{WilsonFlow}
M. L\"uscher,
\emph{Properties and uses of the Wilson flow in lattice QCD},
\emph{JHEP} {\bf 08} (2010) 071.



\bibitem{VolumeEffectsI}
R.~Brower, S.~Chandrasekharan, J.~Negele, U.-J.~Wiese,
\emph{QCD at fixed topology},
\emph{Phys.~Lett.}~{\bf B560} (2003) 64.

\bibitem{VolumeEffectsII}
S.~Aoki, H.~Fukaya, S.~Hashimoto, T.~Onogi,
\emph{Finite-volume QCD at fixed topological charge},
\emph{Phys.~Rev.}~{\bf D76} (2007) 054508.



\bibitem{Wilson}
K.~G.~Wilson, 
\emph{Confinement of quarks},
\emph{Phys.~Rev.}~{\bf D10} (1974) 2445.



\bibitem{Stout}
C.~Morningstar, M.~Peardon,
\emph{Analytic smearing of SU(3) link variables in lattice QCD},
\emph{Phys.~Rev.}~{\bf D69} (2004) 054501.



\bibitem{SommerScale}
R.~Sommer,
\emph{A new way to set the energy scale in lattice gauge theories
and its applications to the static force and $\alpha_s$ in
SU(2) Yang--Mills theory},
\emph{Nucl. Phys.}~{\bf B411} (1994) 839.

\bibitem{GuagnelliEtAl}
M.~Guagnelli, R.~Sommer, H.~Wittig (ALPHA collab.),
\emph{Precision computation of a low-energy reference scale in
quenched lattice QCD},
\emph{Nucl. Phys.}~{\bf B535} (1998) 389.



\bibitem{LambdaParm}
S.~Capitani, M.~L\"uscher, R.~Sommer, H.~Wittig
(ALPHA collab.),
\emph{Non-pertur\-ba\-tive quark mass renormalization in quenched lattice QCD},
\emph{Nucl. Phys.}~{\bf B544} (1999) 669.



\bibitem{TopCharge}
M.~L\"uscher,
\emph{Topology of lattice gauge fields},
\emph{Commun. Math. Phys.}~{\bf 85} (1982) 39.

\bibitem{PhillipsStone}
A.~Phillips, D.~Stone, 
\emph{Lattice gauge fields,
principal bundles and the calculation of the topological charge},
\emph{Commun. Math. Phys.}~{\bf 103} (1986) 599.


\end{thebibliography}
\end{document}